\documentclass[twocolumn,aps, prl,showpacs]{revtex4}

\usepackage{graphics}
\usepackage{graphicx}
\usepackage{dcolumn}
\usepackage{bm}
\usepackage{amsmath,amssymb}

\newcommand{\de}{\partial}

\newcommand{\eq}[2]{\begin{equation} \label{#1} #2 \end{equation}}

\begin{document}

\title{Linear and nonlinear photonic Jackiw-Rebbi states in waveguide arrays}
\author{Truong X. Tran$^{1}$ and Fabio Biancalana$^{2}$}
\affiliation{$^{1}$Dept. of Physics, Le Quy Don University, 236 Hoang Quoc Viet str., 10000 Hanoi, Vietnam\\
$^{2}$School of Engineering and Physical Sciences, Heriot-Watt University, EH14 4AS Edinburgh, UK}
\date{\today}

\begin{abstract}
We study analytically and numerically the optical analogue of the Jackiw-Rebbi states in quantum field theory. These solutions exist at the interface of two binary waveguide arrays which are described by two Dirac equations with opposite sign masses. We show that these special states are topologically robust not only in the linear regime, but also in nonlinear regimes (with both focusing and de-focusing nonlinearity). We also reveal that one can generate the Jackiw-Rebbi states starting from  Dirac solitons.
\end{abstract}
\maketitle

\section{I. INTRODUCTION}
\label{Introduction}

Waveguide arrays (WAs) have been used intensively to simulate the evolution of a non-relativistic quantum mechanical particle in a periodic potential \cite{lederer}. Many fundamental phenomena in non-relativistic classical and quantum mechanics such as Bloch oscillations \cite{pertsch,morandotti2} and Zener tunneling \cite{ghulinyan,trompeter} have been investigated both theoretically and experimentally by using WAs. It was shown in recent studies that most of the nonlinear phenomena usually associated to fiber optics (such as the emission of resonant radiation from solitons and soliton self-frequency shift) can also take place in specially excited WAs, but in the spatial domain rather than in the temporal domain \cite{tranresonant1,tranresonant2}. In addition, a supercontinuum in both frequency and wave number domains can be generated in nonlinear WAs \cite{tranresonant3}. Binary waveguide arrays (BWAs) have been used to mimic relativistic phenomena typical of quantum field theory (QFT), such as Klein tunneling \cite{longhi1,dreisow2}, {\em  Zitterbewegung} (trembling motion of a free Dirac electron) \cite{longhi2,dreisow}, and fermion pair production \cite{longhi3}, which are all based on the properties of the Dirac equation \cite{zeuner}. The discrete gap solitons in BWAs in the {\em classical} context have been investigated both numerically \cite{sukhorukov1,conforti11} and experimentally \cite{morandotti}. Gap and out-gap solitons and breathers in BWAs have been investigated  \cite{johansson1,johansson2}. These gap solitons were already known in \cite{kivshar2} for diatomic lattices, and later derived in their continuum-limit form for the BWA system in \cite{johansson1}. Recently, the explicit suggestion to use BWAs to simulate a quantum nonlinear Dirac equation has been put forward in \cite{trandirac1} where the gap solitons in BWAs have been shown to be connected to Dirac solitons (DSs) in a nonlinear extension of the relativistic one-dimensional (1D) Dirac equation describing the dynamics of a freely moving relativistic particle. The 1D DS stability, its dynamics and different scenarios of soliton interaction have been systematically investigated in \cite{trandirac2}. The formation and dynamics of two-dimensional DSs in square binary waveguide lattices have also been considered in \cite{trandirac3}. The higher-order Dirac solitons in BWAs have been studied in \cite{trandirac4}. Although there is currently no evidence for fundamental quantum nonlinearities, nonlinear versions of the Dirac equation have been studied for a long time, mainly because they allow the exploration of extreme physical environments, such as, for instance, Dirac particles dynamics in astrophysical objects \cite{marini}, or the influence of fermion fields in general relativity singularities \cite{poplawski}. One of the earlier extensions was investigated by Heisenberg himself \cite{Heisenberg57} in the context of field theory and was motivated by the question of mass. In the quantum mechanical context, nonlinear Dirac equations have been used as effective theories in atomic, nuclear and gravitational physics \cite{NLD1,NLD2,NLD3,NLD4}, and in the field of topological insulators \cite{topological}. To this regard, BWAs can offer a unique platform to simulate nonlinear extensions of the Dirac equation when probed at high light intensities.

In this work, we explore the possibility of creating the optical analogue of a special topological state, known in QFT as a {\em Jackiw-Rebbi} (JR) solution \cite{jackiw}. This solution is a type of topological edge state, which exists at the (continuous or discontinuous) interface of two waveguide array regions where the Dirac mass changes sign. We study the linear and {\em nonlinear} properties of the JR-state, which is possible by using the Kerr nonlinearity of the waveguides forming the array. Finally, we explore the excitability of JR-states starting from two Dirac solitons with masses of opposite sign, propagating individually in each waveguide array. This new photonics state can have important applications in light manipulation and circuitry, or in the generation of robust, topologically protected correlated photon sources.

\section{II. GOVERNING EQUATIONS}
\label{geq}

Light propagation in a discrete, periodic binary array of Kerr nonlinear waveguides can be described, in the monochromatic regime, by the following dimensionless coupled-mode equations (CMEs) \cite{longhi1}:
\eq{CWCM}{i\frac{da_{n}(z)}{dz}+\kappa[a_{n+1}(z)+ a_{n-1}(z)] - (-1)^{n} \sigma a_{n} +  \gamma |a_{n}(z)|^{2}a_{n}(z)=0,}
where $a_{n}$ is the electric field amplitude in the $n$th waveguide, $z$ is the longitudinal spatial coordinate, $2\sigma$ and $\kappa$ are the propagation mismatch and the coupling coefficient between two adjacent waveguides of the array, respectively, and $\gamma$ is the nonlinear coefficient of waveguides which is positive for self-focusing, but negative for self-defocusing media. For $n<0$ (for the left-hand side BWA) $\sigma =\sigma_{1}$, whereas for $n\geq0$ (the right-hand side BWA) $\sigma =\sigma_{2}$.

After setting $\Psi_{1}(n) = (-1)^{n}a_{2n}$ and $\Psi_{2}(n) = i(-1)^{n}a_{2n-1}$, and following the standard approach developed in \cite{longhi2,dreisow} we can introduce the continuous transverse coordinate $\xi \leftrightarrow n$ and the  two-component spinor $\Psi(\xi,z)$ = $(\Psi_{1},\Psi_{2})^{T}$ which satisfies the 1D nonlinear Dirac equation \cite{trandirac1}:
\eq{diracequation}{i\de_{z}\Psi = -i\kappa\hat{\sigma}_{x}\de_{\xi}\Psi + \sigma\hat{\sigma}_{z}\Psi - \gamma G,} where the nonlinear terms $G \equiv (|\Psi_{1}|^{2}\Psi_{1},|\Psi_{2}|^{2}\Psi_{2})^{T}$;  $\hat{\sigma}_{x}$ and $\hat{\sigma}_{z}$ are the usual Pauli matrices. Parameter $\sigma$ plays the role of the mass of the Dirac field.

\section{III. LOCALIZED JACKIW-REBBI STATES IN THE LINEAR CASE}
\label{linearcase}

If $\sigma_{1}<0$ and $\sigma_{2}>0$ we get the following exact localized solution of Eq. (\ref{diracequation}) in the linear case:
\eq{solutioncontinuous}{ \Psi(\xi) = \sqrt{\frac{|\sigma_{1}\sigma_{1}|}{\kappa(|\sigma_{1}|+|\sigma_{2}|)}}\left(\begin{array}{cc} 1 \\ i \end{array}\right)e^{-|\sigma(\xi)\xi|/\kappa}.} Even though the solution in the form of Eq. (\ref{solutioncontinuous}) is the exact one to  Eq. (\ref{diracequation}), but it is an approximate solution to the discrete Eq. (\ref{CWCM}). Obviously, this approximation will become better if the beam width gets larger. If $|\sigma_{1}|=|\sigma_{2}|=\sigma_{0}$, one can easily get following exact localized solutions for the discrete Eq. (\ref{CWCM}) without nonlinearity ($\gamma$ = 0) for the following two cases:

If $-\sigma_{1}=\sigma_{2}=\sigma_{0}>0$, one gets:
\eq{solutiondiscrete1}{a_{n} = b_{n}e^{i[\kappa-\sqrt{\sigma_{0}^{2}+\kappa^{2}}]z},} where $b_{n}$ is real and independent of the variable $z$, $b_{2n-1} = b_{2n}$. For $n\geq0$ (the right-hand side BWA) one has the following relationship: $b_{2n}/b_{2n+1} = \alpha \equiv -[\sigma_{0}/\kappa + \sqrt{1+\sigma_{0}^{2}/\kappa^{2}}]$, whereas for $n<0$ (the left-hand side BWA) one has: $b_{2n+1}/b_{2n} = \alpha$.

However, if $\sigma_{1}=-\sigma_{2}=\sigma_{0}>0$, one has:
\eq{solutiondiscrete2}{a_{n} = b_{n}e^{i[\kappa+\sqrt{\sigma_{0}^{2}+\kappa^{2}}]z},} where $b_{n}$ is again real and independent of the variable $z$, $b_{2n-1} = b_{2n}$. For $n\geq0$ one has: $b_{2n}/b_{2n+1} = -\alpha$, whereas for $n<0$ one has: $b_{2n+1}/b_{2n} = -\alpha$.

\begin{figure}[htb]
  \centering \includegraphics[width=0.45\textwidth]{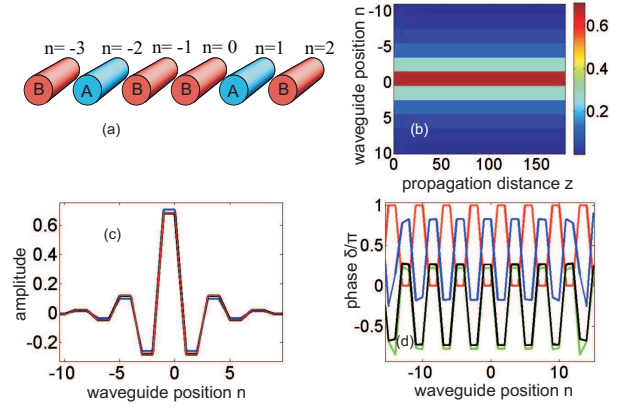}
\caption{\small{(Color online) (a) Illustrative sketch of two BWAs with opposite propagation mismatches located adjacent to each other. (b) Propagation of a beam in the linear regime where Eq. (\ref{solutioncontinuous}) is used as input condition. (c) Amplitudes of the beam at input (blue) and output (black). The red curve represents the solution in the form of Eq.  (\ref{solutiondiscrete1}). (d) Phase pattern of the beam in (a) at different propagation distances $z$ = 0 (red), 100 (blue), 150 (green), and 200 (black). Parameters are $-\sigma_{1} = \sigma_{2} = 1$, $\kappa = 1$, $\gamma = 0$. Two BWAs consist of 841 waveguides in total.}}
  \label{fig1}
\end{figure}

In Fig. \ref{fig1}(a) we show the illustrative sketch of two BWAs with opposite propagation mismatches located adjacent to each other. In Fig. \ref{fig1}(b) we show the propagation of a beam in the linear regime where Eq. (\ref{solutioncontinuous}) is used as input condition for numerically solving Eq. (\ref{CWCM}). Amplitudes of the beam at input (blue) and output (black) are plotted in Fig. \ref{fig1}(c). In Fig. \ref{fig1}(c) we  also plot the red curve representing the exact solution in the form of Eq.  (\ref{solutiondiscrete1}) for the discrete model represented by Eq. (\ref{CWCM}). The red curve and the output black curve in Fig. \ref{fig1}(c) coincide perfectly with each other, therefore the output black curve is hidden behind the red curve. Figure \ref{fig1}(d) shows the phase pattern of the beam at four propagation distances $z$ = 0 (red), 100 (blue), 150 (green), and 200 (black). From Eqs. (\ref{solutioncontinuous}) and (\ref{solutiondiscrete1}), one can easily see that as the waveguide position variable $n$ runs, the phase pattern of the JR states must be periodic as follows: $\delta_{n} = ... (\rho, \rho), (\rho + \pi, \rho + \pi), (\rho, \rho)...$ where $\rho$ also changes with $z$ (fields at two central waveguides n = -1 and 0 are in-phase). These phase patterns are perfectly illustrated in Fig. \ref{fig1}(d). Parameters used for simulations in Fig. \ref{fig1} are $-\sigma_{1} = \sigma_{2} = 1$, $\kappa = 1$, $\gamma = 0$. Two BWAs consist of 841 waveguides in total.

\section{IV. LOCALIZED JACKIW-REBBI STATES IN THE NONLINEAR CASE}
\label{nonlinearcase}

\begin{figure}[htb]
  \centering \includegraphics[width=0.45\textwidth]{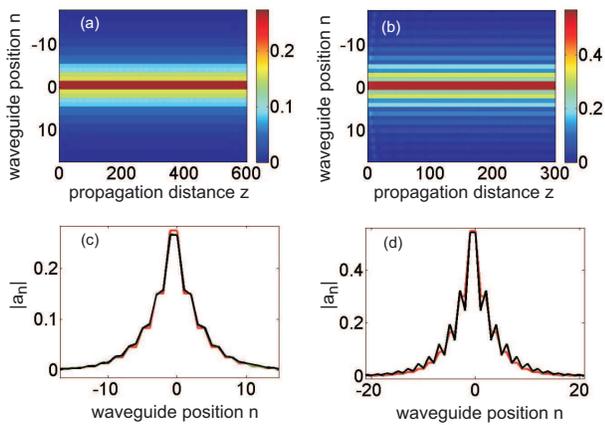}
\caption{\small{(Color online) (a,b) Propagation of a beam with focusing and de-focusing nonlinearity, respectively. (c,d) $|a_{n}|$ of the corresponding beam in (a,b), respectively, at different propagation distances $z$ = 0 (red), 100 (blue), 200 (green), and 300 (black). Parameters: $\gamma$ = 1 and -1 in (a) and (b), respectively; $-\sigma_{1} = \sigma_{2} = 0.6$; $\kappa = 1$. Two BWAs consist of 841 waveguides in total. As input condition for (a,b) Eq. (\ref{solutioncontinuous}) is used, but multiplied by a factor of 0.5 and 1.0 in (a) and (b), respectively.}}
  \label{fig2}
\end{figure}

In Figs. \ref{fig2}(a,b) we show the propagation of a beam with focusing ($\gamma$ = 1) and de-focusing nonlinearity ($\gamma$ = -1), respectively. As input condition for Figs. \ref{fig2}(a,b) Eq. (\ref{solutioncontinuous}) is used, but multiplied by a factor of 0.5 and 1.0 in Figs. \ref{fig2}(a) and \ref{fig2}(b), respectively. Note that the beam widths (FWHM) in Fig. \ref{fig2} are larger than in Fig. \ref{fig1}, because in Fig. \ref{fig2} we use $-\sigma_{1} = \sigma_{2} = 0.6$, whereas $-\sigma_{1} = \sigma_{2} = 1.0$ in Fig. \ref{fig1}, see also Eq. (\ref{solutioncontinuous}). In Fig. \ref{fig2}(c) we plot the beam profile $|a_{n}|$ taken from Fig. \ref{fig2}(a) at four propagation distances $z$ = 0 (red), 100 (blue), 200 (green), and 300 (black). From Figs. \ref{fig2}(a) and \ref{fig2}(c) one can see that the input profile is slightly adjusted to a stable soliton profile which is well conserved during propagation. As a result, three latter curves in Fig. \ref{fig2}(c) are almost identical, and one can only see the black curve, whereas two other curves (blue and green) are hidden behind the black curve. Note that for the localized state in the linear regime illustrated in Fig. \ref{fig1} one has $|a_{2n-1}| = |a_{2n}|$, whereas for the established localized state in the regime with focusing nonlinearity illustrated in Figs. \ref{fig2}(a) and (c), except for the two central sites where $|a_{-1}| = |a_{0}|$, one has $|a_{2n-1}| < |a_{2n}|$ if $n<0$, and  $|a_{2n-1}| > |a_{2n}|$ if $n>0$. Analogously, in Fig. \ref{fig2}(d) we plot the beam profile $|a_{n}|$ taken from Fig. \ref{fig2}(b) at four propagation distances $z$ = 0 (red), 100 (blue), 200 (green), and 300 (black). From Figs. \ref{fig2}(b) and (d) one can see that the input profile is slightly adjusted to a stable soliton profile which is also well conserved during propagation. Note that for the established localized state in the regime with de-focusing nonlinearity illustrated in Fig. \ref{fig2}(b) and (d), except for the two central sites where $|a_{-1}| = |a_{0}|$, one has $|a_{2n-1}| > |a_{2n}|$ if $n<0$, and  $|a_{2n-1}| < |a_{2n}|$ if $n>0$. It is worth mentioning that the phase patterns of two beams in Fig. \ref{fig2} are also identical to the ones illustrated in Fig. \ref{fig1}(d).

\section{V. FORMATION OF JACKIW-REBBI STATES BY DIRAC SOLITONS}
\label{formation}

\begin{figure}[htb]
  \centering \includegraphics[width=0.45\textwidth]{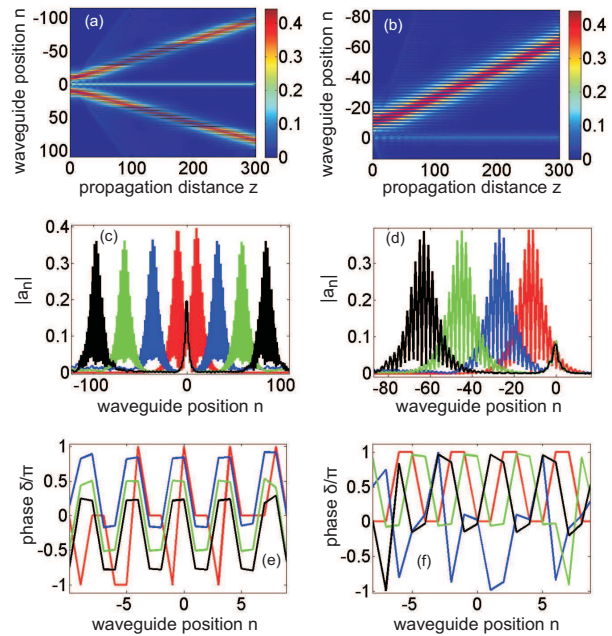}
\caption{\small{(Color online) (a) Formation of a JR state from two out-of-phase Dirac solitons. (b) Formation of a JR state from one Dirac soliton. (c,d) $|a_{n}|$ of the corresponding beam in (a,b), respectively, at different propagation distances $z$ = 0 (red), 100 (blue), 200 (green), and 300 (black). (e,f) Phase patterns of the beams in (c,d), respectively at corresponding propagation distances. Parameters: $\gamma$ = 1, $-\sigma_{1} = \sigma_{2} = 1.0$, $\kappa = 1$.}}
  \label{fig3}
\end{figure}

In Fig. \ref{fig3}(a) we show the formation of a JR state from two out-of-phase Dirac solitons. These Dirac solitons with analytical solutions are taken from Ref. \cite{trandirac1} and are initially motionless in the transverse direction with centers at waveguides $n$ = $\pm$ 10. These two Dirac solitons repel each other during propagation, and at the same time a JR-state is formed at the center of the array. In Figs. \ref{fig3}(c) and \ref{fig3}(e) we plot the beam profiles and its phase patterns, respectively, at four propagation distances $z$ = 0 (red), 100 (blue), 200 (green), and 300 (black) of the JR-state shown in Fig. \ref{fig3}(a). It is worth emphasizing that at the input the overlapping of the two Dirac solitons is weak and we have the profile which is totally different from the one of a JR-state. However, the phase pattern of the field at the center of the array at the input (see the red curve in  Fig. \ref{fig3}(e)) is identical to the one of the JR-state as illustrated in Fig. \ref{fig1}(d) (note the trivial fact that the phase difference equal to 2$\pi$ means fields are just in-phase). This requirement for phase pattern at the input is crucial for formation of the JR-states later during propagation. For two symmetric Dirac solitons located at two BWAs as shown in Fig. \ref{fig3}(a) this phase pattern is only obtained if these two Dirac solitons are initially out-of-phase. If this condition is satisfied, then a JR-state with right beam profile and phase pattern will be formed as clearly shown in Figs. \ref{fig3}(a), \ref{fig3}(c), and \ref{fig3}(e). Of course, the closer the initial Dirac solitons, the larger the amplitude of the established JR-state.

We have shown in Figs. \ref{fig3}(a), \ref{fig3}(c), and \ref{fig3}(e) that a JR-state can be formed from two out-of-phase Dirac solitons. In the same way, we show in Figs. \ref{fig3}(b), \ref{fig3}(d), and \ref{fig3}(f) that a \emph{quasi} JR-state can also be formed from just one Dirac soliton which is initially located close to the border of the two BWAs. However, in this case, the peak amplitude of the established \emph{quasi} JR-state as shown in Fig. \ref{fig3}(d) is about three times weaker than the JR-state shown in Fig. \ref{fig3}(c). The phase pattern of this beam as illustrated in Fig. \ref{fig3}(f) is also not perfectly identical to the one of a true JR-state.

\section{VI. CONCLUSIONS}
\label{conclusion}

We have analytically and numerically demonstrated the existence of the optical analogue of edge states - known in quantum field theory as the Jackiw-Rebbi states - formed at the interface of two BWAs having propagation mismatches with opposite signs. Remarkably, the localized JR states can be formed both in linear and nonlinear regimes and can be divided into two types with different amplitude profiles depending on the signs of propagation mismatches at the interface. We have also shown the excitation of JR-states from Dirac solitons. Due to the robustness of the topological JR-states, we envision important future applications in
the generation of topologically-protected correlated photons, using a similar scheme as proposed recently in Ref. \cite{blanco}.

\section{ACKNOWLEDGMENT}
This work is supported by the Vietnam National Foundation for Science and Technology (NAFOSTED) under grant number 103.03-2016.01.

\end{document}